\pdfoutput=1

\documentclass{comjnlarxiv}
\usepackage{flafter}
\usepackage{float}

\usepackage{flushend}
\usepackage{amsmath}
\usepackage{listings}
\usepackage{booktabs}
\newcommand{\microop}{$\mu$op}
\newcommand{\microops}{\microop{}s}
 \DeclareMathOperator{\popcount}{popcount}
\DeclareMathOperator{\bitwiseand}{AND}
\DeclareMathOperator{\bitwiseor}{OR}
\DeclareMathOperator{\bitwisexor}{XOR}

\usepackage{microtype}
\usepackage{inconsolata}
\usepackage{tikz}

\usetikzlibrary{chains,arrows,automata,decorations.markings,positioning,calc,decorations.pathreplacing,patterns}
\usepackage{array}
\newcolumntype{x}[1]{>{\centering\arraybackslash\hspace{0pt}}p{#1}}
\lstdefinestyle{customc}{%
  belowcaptionskip=1\baselineskip,
  breaklines=true,
  xleftmargin=\parindent,
  language=C,
  showstringspaces=false,
  basicstyle=\footnotesize\ttfamily,
  keywordstyle=\bfseries\color{green!40!black},
  numberstyle=\tiny,
  commentstyle=\itshape\color{purple!40!black},
  identifierstyle=\bfseries\color{black},
  stringstyle=\color{orange},
   morekeywords={uint64_t,uint32_t,__m256i,__m128i,UINT64_C},
}

\usepackage[english]{babel}
\usepackage[utf8x]{inputenc}
\usepackage{amsmath}
\usepackage{url}
\usepackage{underscore}
\usepackage{siunitx}
\usepackage[colorinlistoftodos]{todonotes}
\usepackage{algorithm}
\usepackage[noend]{algorithmic}

\usepackage{graphicx}

\begin{document}

\title[Faster Population Counts Using AVX2 Instructions]{Faster Population Counts Using AVX2 Instructions}

\shortauthors{W. Muła, N. Kurz and D. Lemire}

\author{Wojciech Muła} 
\email{wojciech_mula@poczta.onet.pl}

\author{Nathan Kurz}
\email{natekurz@gmail.com}

\author{Daniel Lemire}
\affiliation{$^\star$Universit\'e du Qu\'ebec (TELUQ), Canada}
\email{lemire@gmail.com}
\address{5800 Saint-Denis, Montreal (Quebec)
H2S 3L5 Canada}



\begin{abstract}
Counting the number of ones in a binary stream is a common
operation in database, information-retrieval, cryptographic  and machine-learning applications.
Most processors have dedicated instructions to count the number of ones in a word (e.g., \texttt{popcnt} on x64 processors). Maybe surprisingly, we show that a vectorized approach using SIMD instructions
can be twice as fast as using the dedicated instructions on recent Intel processors.
The benefits can be even greater for applications such as similarity
measures (e.g., the Jaccard index) that require additional Boolean
operations.
 Our approach has been adopted by LLVM:  it is used by its popular C compiler (Clang).
\end{abstract}
\lstset{escapechar=@,style=customc}
\maketitle
\section{Introduction}
%

We can represent all sets of integers in $\{0,1,\ldots, 63\}$ using
a single 64-bit word.  For example, the word \texttt{0xAA} (\texttt{0b10101010})  represents the set $\{1,3,5,7\}$. Intersections and unions between
such sets
can be computed using a single bitwise logical operation on each pair
of words
(AND, OR).
 We can generalize this idea to sets of integers in $\{0,1,\ldots, n-1 \}$  using $\lceil n/64 \rceil $~64-bit words.
We call such data structures  \emph{bitsets};  they are also known as a bit vectors, bit arrays or bitmaps.
Bitsets are ubiquitous in software, found in databases~\cite{SPE:SPE2402}, version control systems~\cite{gitewah}, search engines~\cite{SPE:SPE2325,SPE:SPE2326,RoaringDocIdSetBlogPost}, and so forth. Languages such as Java and C++
come with their own bitset classes (\texttt{java.util.BitSet}
and \texttt{std::bitset} respectively).

The cardinality of a bitset (the number of one bits, each
representing an element in the set) is commonly called a population
count, a popcount, a Hamming weight, a sideways addition, or sideways
sum. For example,
 the population counts of the words \texttt{0xFFFF}, \texttt{0xAA} and \texttt{0x00}  are 16, 4 and 0 respectively.
 A frequent purpose for the population count is to determine the size of the intersection or union between two
 bitsets. In such cases, we must first apply a logical operation on pairs of
 words (AND, OR) and then compute the population count of the resulting words.
 For example, the cardinality of the intersection of the sets $A=\{4,5,6,7\}$ and $B=\{1,3,5,7\}$
represented by the words \texttt{0xF0} and \texttt{0xAA} can be computed as
$\vert A \cap B \vert = \popcount(\texttt{0xF0} \bitwiseand \texttt{0xAA})=\popcount(\texttt{0xA0})=2$.

 Population-count functions are used in cryptography~\cite{hilewitz2004comparing}, e.g., as part of randomness tests~\cite{suciu2011never} or to generate pseudo-random permutations~\cite{stefanov2012fastprp}.
They can help find duplicated web pages~\cite{Manku:2007:DNW:1242572.1242592}.
  They are frequently used in bioinformatics~\cite{prokopenko2016utilizing,Lacour2015,li2012wham}, ecology~\cite{dambros2015effects}, chemistry~\cite{zhang2015design},  and so forth.
Gueron and Krasnov use population-count instructions as part of a fast sorting algorithm~\cite{Gueron01012016}.

The computation of the population count is so important that
commodity processors have dedicated instructions: \texttt{popcnt} for x64 processors and \texttt{cnt} for the 64-bit ARM architecture.\footnote{The x64 \texttt{popcnt} instruction was first available in the  Nehalem microarchitecture, announced in 2007 and released in November 2008. The ARM \texttt{cnt} instruction was released as part of the Cortex-A8 microarchitecture, published in 2006~\cite{7477864}.} The
 x64 \texttt{popcnt} instruction is fast: on recent Intel processors, it has a throughput of
 one instruction per cycle~\cite{fog2016instruction} (it can
execute once per cycle) and a latency of 3 cycles (meaning that the
result is available for use on the third cycle after execution). It is available in common C and C++ compilers as the intrinsic
\texttt{_mm_popcnt_u64}. In Java, it is available as the \texttt{Long.bitCount} intrinsic.

Commodity
PC processors also support Single-Instruction-Multiple-Data (SIMD) instructions.  Starting with the Haswell microarchitecture (2013), Intel processors support the AVX2 instruction set which offers  rich support for 256-bit vector registers.
The contest between a dedicated instruction operating on 64-bits at a
time (\texttt{popcnt}) and a series of vector instructions operating on
256-bits at a time (AVX2) turns out to be interesting.
In fact, we show that we can achieve twice the speed of the an optimized \texttt{popcnt}-based  function using AVX2: 0.52 versus 1.02~cycles per 8~bytes on large arrays.
Our claim has been thoroughly validated:
at least one major C compiler (LLVM's Clang) uses our technique~\cite{clangMula2015}.

Thus, in several instances, SIMD instructions might be preferable to
dedicated non-SIMD instructions if we are just interested in the population count of a bitset. But what if we seek the cardinality of the intersection or union, or simply the Jaccard index between two bitsets?
Again, the AVX2 instructions prove useful, more than doubling the
speed ($2.4\times$) of the computation against an optimized function using the  \texttt{popcnt} instruction.

\section{Existing Algorithms and Related Work}

Conceptually, one could compute the population count by checking the value of each bit individually by calling \texttt{count += (word >> i) \& 1} for \texttt{i} ranging from 0 to 63, given that \texttt{word} is a 64-bit word.  While this approach scales linearly in the number of input words, we expect it to be slow since it requires multiple operations for each bit in each word. It is $O(n)$---the best that we can do---but with a high constant factor.

\begin{figure}[h]
\centering
\begin{tikzpicture}[level distance=1.5cm,
  level 1/.style={sibling distance=3cm},
  level 2/.style={sibling distance=1.5cm}]
  \node[fill=red!10,draw] {population count of 4~bits}
    child {node[fill=blue!10,draw] {sum of bits 1 \& 2}
      child {node[fill=yellow!30,draw]  {1$^{\text{st}}$ bit}  edge from parent[<-,thick,>=latex]}
      child {node[fill=yellow!30,draw] {2$^{\text{nd}}$ bit} edge from parent[<-,thick,>=latex]}
      edge from parent[<-,thick,>=latex]
    }
    child {node[fill=blue!10,draw] {sum of bits 3 \& 4}
    child {node[fill=yellow!30,draw] {3$^{\text{rd}}$ bit} edge from parent[<-,thick,>=latex]}
      child {node[fill=yellow!30,draw] {4$^{\text{th}}$ bit} edge from parent[<-,thick,>=latex]}
      edge from parent[<-,thick,>=latex]
    };
\end{tikzpicture}
\caption{\label{fig:tree-of-adders} A tree of adders to compute the population count of four bits in two steps.}\end{figure}
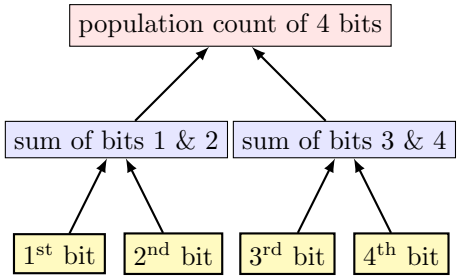

Instead, we should prefer approaches with fewer operations per word.
We can achieve the desired result with a tree of adders and bit-level parallelism.
In  Fig.~\ref{fig:tree-of-adders}, we illustrate the idea over words of 4~bits (for simplicity).
We implement this approach with two lines of code.
\begin{enumerate}
\item We can sum the individual bits to 2-bit subwords
with the line
of C code: \texttt{(~x~\&~0b0101~) + (~(~x~>>~1~)~\&~0b0101~)}. This takes us from the bottom of the tree to the second level. We say to this step exhibits \emph{bit-level parallelism} since two sums are executed at once, within the same 4-bit word.
\item We can then sum the values stored in the 2-bit subwords into a single 4-bit subword with another line of C code:  \texttt{(~x~\&~0b0011~) + (~(~x~>>~2~)~\&~0b0011~)}.
\end{enumerate}
Fig.~\ref{fig:tradd} illustrates a non-optimized (naive) function that computes the population count of a 64-bit word in this manner.

\begin{figure}[h]\centering
\begin{tabular}{c}
\begin{lstlisting}
uint64_t c1  = UINT64_C(0x5555555555555555);
uint64_t c2  = UINT64_C(0x3333333333333333);
uint64_t c4  = UINT64_C(0x0F0F0F0F0F0F0F0F);
uint64_t c8  = UINT64_C(0x00FF00FF00FF00FF);
uint64_t c16 = UINT64_C(0x0000FFFF0000FFFF);
uint64_t c32 = UINT64_C(0x00000000FFFFFFFF);

uint64_t count(uint64_t x) {
  x = (x & c1) + ((x >> 1) & c1);
  x = (x & c2) + ((x >> 2) & c2);
  x = (x & c4) + ((x >> 4) & c4);
  x = (x & c8) + ((x >>  8)  & c8);
  x = (x & c16)+ ((x >> 16)) & c16);
  return (x & c32) + ((x >> 32) & c32);
}

\end{lstlisting}
\end{tabular}
\caption{\label{fig:tradd}A naive tree-of-adders  function in C}
\end{figure}

\begin{figure}[h]\centering
\begin{tabular}{c}
\begin{lstlisting}[style=customc]
uint64_t c1  = UINT64_C(0x5555555555555555);
uint64_t c2  = UINT64_C(0x3333333333333333);
uint64_t c4  = UINT64_C(0x0F0F0F0F0F0F0F0F);

uint64_t count(uint64_t x) {
  x -= (x >> 1) & c1;
  x = (( x >> 2) & c2) + (x & c2);
  x = ( x  + (x >> 4) ) & c4;
  x *= UINT64_C(0x0101010101010101);
  return x >> 56;
}
\end{lstlisting}
\end{tabular}
\caption{\label{fig:wwg}The Wilkes-Wheeler-Gill  function in C}
\end{figure}

A fast and  widely used tree-of-adder function to compute the population count has been attributed by Knuth~\cite{KnuthV4} to a 1957 textbook by Wilkes, Wheeler and Gill~\cite{Wilkes}: see Fig.~\ref{fig:wwg}. It involves far fewer than 64~instructions and we expect it to be several times faster than a naive function checking the values of each bit and faster than the naive tree-of-adder approach on
processor with a sufficiently fast 64-bit integer multiplication (which includes all x64 processors).
\begin{itemize}
\item The first two lines in the \texttt{count} function  correspond to the first two levels of our simplified tree-of-adders \texttt{count} function  from Fig.~\ref{fig:tree-of-adders}. The first line has been optimized. We can verify the optimization by checking that for each possible 2-bit word, we get the sum of the bit values:
\begin{itemize}
\item \texttt{0b11~-~0b01~= 0b10~=~2},
\item \texttt{0b10~-~0b01~=~0b01 =~1},
\item \texttt{0b01~-~0b00~=~0b01 =~1},
\item \texttt{0b00~-~0b00~=~0b00 =~0}.
\end{itemize}
\item After the first two lines, we have 4-bit population counts (in $\{0b0000,0b0001,0b0010,0b0011,0b0100\}$) stored in 4-bit subwords. The next line sums consecutive 4-bit subwords to bytes. We use the fact that the most significant bit of each 4-bit subword is zero.
\item The multiplication and final shift sum all bytes in an efficient way.
Multiplying \texttt{x} by \texttt{0x0101010101010101} is equivalent to
summing up  \texttt{x}, \texttt{x << 8}, \texttt{x << 16}, \ldots, \texttt{x << 56}. The total population count is less than 64, so that the sum of all bytes from  \texttt{x} fits in a single byte value (in $[0,256)$).  In that case, the most significant 8 bits from the product  is the sum of all eight byte values.
\end{itemize}

Knuth also attributes  another common technique to Wegner~\cite{Wegner} (see Fig.~\ref{fig:wegner}) that could be competitive when the population count is relatively low (e.g., less than 4 one bit per 64-bit word). When the population count is expected to be high (e.g., more than 60 one bit per 64-bit words), one could simply negate the words prior to using the function so as to count the number of zeros instead. The core insight behind the Wegner function is that the line of C code \texttt{x \&= x - 1} sets to zero the least significant bit of \texttt{x}, as one can readily check. On an x64 processor,
the expression \texttt{x \&= x - 1} might be compiled to the
\texttt{blsr} (reset lowest set bit) instruction.
On current generation processors, this instruction achieves a
throughput of two instructions per cycle with a  latency of one cycle~\cite{fog2016instruction}.
The downside of the Wegner approach for modern processors is that the
unpredictable loop termination adds a mispredicted branch
penalty of at least 10~cycles~\cite{Rohou:2015:BPP:2738600.2738614}, which for short loops can be more expensive than the
operations performed by the loop.

\begin{figure}[hbt]\centering
\begin{tabular}{c}
\begin{lstlisting}
int count(uint64_t x) {
  int v = 0;
  while(x != 0) {
    x &= x - 1;
    v++;
  }
  return v;
}
\end{lstlisting}
\end{tabular}
\caption{\label{fig:wegner}The Wegner  function in C. }
\end{figure}

Another simple and common technique is based on tabulation.
For example, one might create a table that contains the corresponding
population count for each possible byte value, and then look up and
sum the count for each byte. Such a table would require only 256~bytes. A population count for a 64-bit word would require only eight table look-ups and seven additions. On more powerful processor, with more cache, it might be beneficial to create a larger table, such as one that has a population count for each possible \texttt{short} value (2 bytes) using 64\,KB. Each doubling of the bit-width covered by the table halves the number of table lookups, but squares the memory required for the table.

We can improve the efficiency of tree-of-adders techniques by \emph{merging} the trees across words~\cite{Lauradoux}.
To gain an intuition for this approach,
consider that in the
Wilkes-Wheeler-Gill approach, we use
4-bit subwords to store the population count of four consecutive bits. Such a population count takes a value in $\{0,1,2,3,4\}$, yet a 4-bit integer can represent all integers in $[0,16)$.
Thus, as a simple optimization, we could
accumulate the 4-bit counts across three different words instead of a single one.
Next consider that if you sum two 4-bit subwords (representing integers in $[0,16)$) the result is in $[0,32)$ whereas an 8-bit subword (a byte) can represent all integers in $[0,256)$, a range that is four times larger. Hence, we can accumulate the counts over four triple of words. These two optimizations combined lead to a function to compute the population count of twelve words at once (see Fig.~\ref{fig:lauradoux}) faster than would be possible if we processed each word individually.

\begin{figure}[h]\centering
\begin{tabular}{c}
\begin{lstlisting}
uint64_t count(uint64_t *input) {
 uint64_t m1 = UINT64_C(0x5555555555555555);
 uint64_t m2 = UINT64_C(0x3333333333333333);
 uint64_t m4 = UINT64_C(0x0F0F0F0F0F0F0F0F);
 uint64_t m8 = UINT64_C(0x00FF00FF00FF00FF);
 uint64_t m16= UINT64_C(0x0000FFFF0000FFFF);
 uint64_t acc = 0;
 for (int j = 0; j < 12; j += 3) {
    uint64_t count1  =  input[j + 0];
    uint64_t count2  =  input[j + 1];
    uint64_t half1   =  input[j + 2];
    uint64_t half2   =  input[j + 2];
    half1  &=  m1;
    half2   = (half2  >> 1) & m1;
    count1 -= (count1 >> 1) & m1;
    count2 -= (count2 >> 1) & m1;
    count1 +=  half1;
    count2 +=  half2;
    count1  = (count1 & m2)
              + ((count1 >> 2) & m2);
    count1 += (count2 & m2)
              + ((count2 >> 2) & m2);
    acc  += (count1 & m4)
              + ((count1 >> 4) & m4);
 }
 acc = (acc & m8) + ((acc >>  8)  & m8);
 acc = (acc       +  (acc >> 16)) & m16;
 acc =  acc       +  (acc >> 32);
 return acc;
}
\end{lstlisting}
\end{tabular}
\caption{\label{fig:lauradoux}The Lauradoux population count in C for sets of 12~words. }
\end{figure}

However, even before Lauradoux proposed this improved function, Warren~\cite{warren2007} had presented a superior alternative attributed to a newsgroup posting from 1997 by Seal, inspired from earlier work by Harley. This approach, henceforth called Harley-Seal,  is based on a  carry-save adder (CSA). Suppose you are given three~bit values ($a,b,c\in \{0,1\}$) and you want to compute their sum ($a+b+c\in \{0,1,2,3\}$). Such a sum fits in a 2-bit word. The value of the least significant bit is given by $(a\bitwisexor b) \bitwisexor c$ whereas the most significant bit is given by $(a \bitwiseand b) \bitwiseor ( (a \bitwisexor b) \bitwiseand c )$.
Table~\ref{table:sum} illustrates these expressions: the least significant bit ($(a\bitwisexor b) \bitwisexor c$) takes value 1 only when $a+b+c$ is odd and the most significant bit takes value 1 only when two or three of the input bits ($a,b,c$) are set to 1.
There are many possible expressions to compute the most significant bit, but the chosen expression is convenient because it reuses the $a \bitwisexor b$ expression from the computation of the least significant bit. Thus, we can sum three~bit values to a 2-bit counter using 5~logical operations.
We can generalize this approach to work on all 64-bits
in parallel.  Starting with three~64-bit input words, we can generate
two new output words: $h$, which holds the 64~most significant bits, and $l$, which contains the corresponding 64~least significant bits.
We effectively compute 64~sums in parallel using bit-level parallelism. Fig.~\ref{fig:csa} presents an efficient implementation in C of this idea. The function uses 5~bitwise logical operations (two XORs, two ANDs and one OR): it is optimal with respect to the number of such operations~\cite[7.1.2]{KnuthV4A}. However, it requires at least three~cycles to complete due to data dependencies.

\begin{figure}[h]\centering
\begin{tabular}{c}
\begin{lstlisting}
void CSA(uint64_t* h, uint64_t* l,
    uint64_t a, uint64_t b, uint64_t c) {
  uint64_t u = a ^ b;
  *h = (a & b) | (u & c);
  *l = u ^ c;
}
\end{lstlisting}
\end{tabular}
\caption{\label{fig:csa}A C function  implementing a bitwise parallel carry-save adder (CSA). Given three input words $a,b,c$, it generates two new words $h,l$ in which each bit represents the high and
low bits in the bitwise sum of the bits from $a$, $b$, and $c$. }
\end{figure}

\begin{table}
\caption{Sum of three bits $a+b+c$. We use $\oplus$ for XOR, $\land$ for AND and $\lor$ for OR.\label{table:sum}}\centering\small
\vspace{0.5cm}
\begin{tabular}{ccc|c|x{1.5cm}x{1.5cm}}
\toprule
$a$ & $b$ & $c$ & $a+b+c$ &  $(a\oplus b) \oplus  c$  & $(a \land b)  \lor  \allowbreak ( (a \oplus  b) \land c )$
\\\midrule
0 & 0 & 0 & 0 & 0 & 0\\
0 & 0 & 1 & 1 & 1 & 0\\
0 & 1 & 0 & 1 & 1 & 0\\
1 & 0 & 0 & 1 & 1 & 0\\
0 & 1 & 1 & 2 & 0 & 1\\
1 & 0 & 1 & 2 & 0 & 1\\
1 & 1 & 0 & 2 & 0 & 1\\
1 & 1 & 1 & 3 & 1 & 1\\
\bottomrule
\end{tabular}
\end{table}

\begin{figure}
\centering
\begin{tikzpicture}[thick,scale=0.9, every node/.style={transform shape},node distance=0cm,start chain=1 going right,
start chain=9 going right,start chain=10 going right,start chain=11 going right,start chain=12 going right,start chain=13 going right]
 \tikzstyle{mytape}=[minimum height=0.7cm,minimum width=1cm]

    \node(A3) [on chain=9,mytape] {\ldots};
    \node(A4) [on chain=9,mytape] {\parbox{1cm}{\centering \texttt{twos}\\(input)}};
    \node(A5) [on chain=9,mytape] {\parbox{1cm}{\centering \texttt{ones}\\(input)}};
    \node(A6) [on chain=9,mytape] {\parbox{1cm}{\centering $d_i$\\(input)}};
    \node(A7) [on chain=9,mytape]
    {\parbox{1cm}{\centering $d_{i+1}$\\(input)}};

\node(csa1)[on chain=10,fill=gray!30,minimum width=3cm,below=0.5cm of A6] {CSA};
    \node(A8) [on chain=10,mytape]
    {\parbox{1cm}{\centering $d_{i+2}$\\(input)}};
    \node(A9) [on chain=10,mytape]
    {\parbox{1cm}{\centering $d_{i+3}$\\(input)}};
\draw[->,blue, >=latex,thick] (A5.south)  to    (A5.south |- csa1.north);
\draw[->,blue, >=latex,thick] (A6.south)  to    (A6.south |- csa1.north);
\draw[->,blue, >=latex,thick] (A7.south)  to    (A7.south |- csa1.north);
\node(csa2)[on chain=11,fill=gray!30,minimum width=3cm,below=0.5cm of A8] {CSA};
   \node(A10) [on chain=11,mytape] {\ldots};

\draw[->, blue,>=latex,thick]
    ($(csa1.south east)!0.22!(csa1.south)$) to[out=270,in=90]
    ($(csa2.north west)!0.35!(csa2.north)$);
\draw[->, blue,>=latex,thick] (A8.south)  to    (A8.south |- csa2.north);
\draw[->, blue,>=latex,thick] (A9.south)  to    (A9.south |- csa2.north);

\node(csa3)[on chain=12,below=1cm of A10,minimum width=4cm] {};
\node(csa12)[fill=gray!30,minimum width=3cm,left=0.5cm of csa3] {CSA};

\draw[->,red, >=latex,thick]
    (A4.south) to[out=270,in=90]
    ($(csa12.north west)!0.25!(csa12.north)$);

\node(foursoutput)  [on chain=13,mytape,below=1cm of $(csa12.south west)!0.35!(csa12.south)$,minimum width=2cm]
{\parbox{1cm}{\centering \texttt{fours}\\(output)}};

\draw[->,>=latex,thick]
    ($(csa12.south west)!0.35!(csa12.south)$) to[out=270,in=90]
    (foursoutput.north);

\node(twosoutput)  [minimum width=2cm,below=1cm of $(csa12.south east)!0.35!(csa12.south)$]
{\parbox{1cm}{\centering \texttt{twos}\\(output)}};

\draw[->,red, >=latex,thick]
    ($(csa12.south east)!0.35!(csa12.south)$) to[out=270,in=90]
    (twosoutput.north);

\node(onesoutput)  [mytape,minimum width=2cm,below=1.5cm of $(csa2.south east)!0.35!(csa2.south)$]
{\parbox{1cm}{\centering \texttt{ones}\\(output)}};

\draw[->, blue,>=latex,thick]
    ($(csa2.south east)!0.35!(csa2.south)$) to[out=270,in=90]
    (onesoutput.north);

\draw[->,red, >=latex,thick]
    ($(csa2.south west)!0.35!(csa2.south)$) to[out=270,in=90]
    ($(csa12.north east)!0.35!(csa12.north)$);

\draw[->,red, >=latex,thick]
    ($(csa1.south west)!0.35!(csa1.south)$) to[out=270,in=90]
    (csa12.north);
\end{tikzpicture}
\caption{\label{fig:hsillustration}Harley-Seal algorithm aggregating four new inputs ($d_i, d_{i+1}, d_{i+2}$, $d_{i+3}$) to inputs \texttt{ones} and \texttt{twos}, producing new values of \texttt{ones}, \texttt{twos} and \texttt{fours}. }
\end{figure}
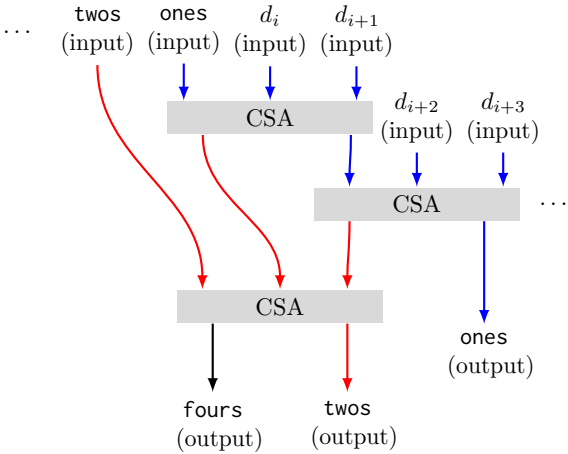

From such a CSA function, we can derive an efficient population count. Suppose we start with three words serving as counters (initialized at zero): one for the least significant bits (henceforth \texttt{ones}),
another one for the second least significant bits (\texttt{twos}, so named
because each bit set  represents 2~input bits), and another for
the third least significant bits (\texttt{fours}, representing 4~input bits).
We can proceed as follows; the first few steps are illustrated in Fig.~\ref{fig:hsillustration}. We start with a word serving as a population counter $c$ (initialized at zero).  Assume with we have a number of words $d_1,d_2,\ldots$ divisible by 8. Start with $i=0$.
\begin{itemize}
\item Load two new words ($d_i,d_{i+1}$). Use the CSA function to sum \texttt{ones}, $d_i$ and $d_{i+1}$, write the least significant bit of the sum to \texttt{ones} and store the carry bits in a temporary register (noted \texttt{twosA}). We repeat with the next two input words. Load $d_{i+2},d_{i+3}$, use the CSA function to sum \texttt{ones}, $d_i$ and $d_{a+i}$, write the least significant bit of the sum to \texttt{ones} and store the carry bits in a temporary register (noted \texttt{twosB}).
\item At this point, we have three words containing second least significant bits (\texttt{twos}, \texttt{twosA}, \texttt{twosB}). We sum them up using a CSA, writing back the result to \texttt{twos} and the carry bits to a temporary register \texttt{foursA}.
\item We do with $d_{i+4},d_{i+5}$ and $d_{i+6},d_{i+7}$ as we did with $d_i,d_{i+1}$ and $d_{i+2},d_{i+3}$. Again we have  three words containing second least significant bits (\texttt{twos}, \texttt{twosA}, \texttt{twosB}). We sum them up with CSA, writing the result to \texttt{twos} and to a carry-bit temporary register  \texttt{foursB}.
\item At this point, we have three words containing third least significant bits (\texttt{fours}, \texttt{foursA}, \texttt{foursB}). We can sum them up with a CSA, write the result back to \texttt{fours}, storing the carry bits in a temporary register \texttt{eights}.
\item We compute the population count of the word  \texttt{eights} (e.g, using the Wilkes-Wheeler-Gill population count) and increment the counter $c$ by the population count.
\item Increment $i$ by 8 and continue for as long as we have new words.
\end{itemize}
When the algorithm terminates, multiply $c$ by 8. Compute the population count of \texttt{fours}, multiply the result by 4 and add to $c$. Do similarly with \texttt{twos} and \texttt{ones}. The counter $c$ contains the population count. If the number of input words is not divisible by 8, adjust accordingly with the leftover words (e.g, using the Wilkes-Wheeler-Gill population count).

In that particular implementation of this idea, we used blocks of eight words. More generally, the Harley-Seal approach works with blocks of $2^n$~words for $n=3,4,5, \ldots$ (8, 16, 32, \ldots).
 We need $2^n-1$ CSA function calls when using $2^n$~words, and one call to an auxiliary function  (e.g., Wilkes-Wheeler-Gill). If we expect the auxiliary function to be significantly more expensive than the CSA function calls, then larger blocks should lead to higher performance, as long as we have enough input data and many available registers. In practice, we found that using blocks of sixteen~words works well on current processors (see Fig.~\ref{fig:harleyseal16}). This approach is only worthwhile if we have at least 16~input words (64-bits/word~$\times$~16~words~$=128$~bytes).

\begin{figure}[h]\centering
\begin{tabular}{c}
\begin{lstlisting}
uint64_t harley_seal(uint64_t * d,
      size_t size) {
 uint64_t total = 0, ones = 0, twos = 0,
    fours = 0, eights = 0, sixteens = 0;
 uint64_t twosA, twosB, foursA, foursB, eightsA, eightsB;
 for(size_t i = 0; i < size - size % 16;
    i += 16) {
  CSA(&twosA, &ones, ones, d[i+0], d[i+1]);
  CSA(&twosB, &ones, ones, d[i+2], d[i+3]);
  CSA(&foursA, &twos, twos, twosA, twosB);
  CSA(&twosA, &ones, ones, d[i+4], d[i+5]);
  CSA(&twosB, &ones, ones, d[i+6], d[i+7]);
  CSA(&foursB, &twos, twos, twosA, twosB);
  CSA(&eightsA, &fours, fours, foursA, foursB);
  CSA(&twosA, &ones, ones, d[i+8], d[i+9]);
  CSA(&twosB, &ones, ones, d[i+10],d[i+11]);
  CSA(&foursA, &twos, twos, twosA, twosB);
  CSA(&twosA, &ones, ones, d[i+12],d[i+13]);
  CSA(&twosB, &ones, ones, d[i+14],d[i+15]);
  CSA(&foursB, &twos, twos, twosA, twosB);
  CSA(&eightsB, &fours, fours, foursA,
     foursB);
  CSA(&sixteens, &eights, eights, eightsA,
     eightsB);
  total += count(sixteens);
 }
 total = 16 * total + 8 * count(eights)
      + 4 * count(fours) + 2 * count(twos)
      + count(ones);
 for(size_t i = size - size % 16 ; i < size; i++)
  total += count(d[i]);
 return total;
}
\end{lstlisting}
\end{tabular}
\caption{\label{fig:harleyseal16}A C function implementing the Harley-Seal population count over an array of 64-bit words. The \texttt{count} function could be the Wilkes-Wheeler-Gill function. }
\end{figure}

The functions we presented thus far still have their uses when programming with high-level languages without convenient access to dedicated functions (e.g., JavaScript, Go) or on limited hardware. However, they are otherwise obsolete when a sufficiently fast  instruction is available, as is the case on recent x64 processors with \texttt{popcnt}. The  \texttt{popcnt} instruction has a reciprocal throughput\footnote{The reciprocal throughput is the number of processor clocks it takes for an instruction to execute.} of one instruction per cycle.
With a properly constructed loop, the load-\texttt{popcnt}-add sequence can be executed in a single cycle, allowing for a population  count function that processes 64-bits per cycle.

\subsection{Existing Vectorized Algorithms}

To our knowledge, the  first published vectorized population count on Intel processor was proposed by  Muła in  2008~\cite{Mula2008}.
It is  a vectorized form of tabulation on 4-bit subwords.
Its key ingredient is the SSSE3 vector instruction \texttt{pshufb} (see Table~\ref{ref:sseinstructions}).
The \texttt{pshufb}  instruction shuffles the
input bytes  into a new vector containing the same byte values in a
(potentially) different order.
It takes an input register $v$ and a control mask $m$, treating both as vectors of sixteen~bytes.
 Starting from $v_0, v_1, \ldots, v_{16}$, it outputs a new vector $(v_{m_0},v_{m_1},v_{m_2},v_{m_3}, \ldots, v_{m_{15}})$ (assuming that $0 \leq m_i < 16$ for $i=0,1,\ldots, 15$).
 Thus, for example, if the mask $m$ is $0,1,2,\ldots, 15$, then we have the identify function. If the mask $m$ is $15, 14, \ldots, 0$, then the byte order is reversed. Bytes are allowed to be repeated in
the output vector, thus the mask $0,0,\ldots, 0$ would produce a vector containing only the first input byte, repeated sixteen times.
  It is a fast instruction with a reciprocal throughput and latency of one cycle on current Intel processors, yet it effectively ``looks up'' 16 values at once.
 In our case, we use a fixed input register made of the input bytes $0,1,1,2,1,2,2,3,1,2,2,3,2,3,3,4$
 corresponding to the population counts of all possible 4-bit integers $0,1,2,3,\ldots, 15$.
 Given an array of sixteen bytes, we can call \texttt{pshufb} once, after selecting the least  significant 4~bits of each byte (using a bitwise AND) to gather sixteen population counts on sixteen 4-bit subwords. Next, we right shift by four bits each byte value, and call \texttt{pshufb}  again to gather  sixteen counts of the most significant 4~bits of each byte.
 We can sum the two results to obtain sixteen population counts, each
corresponding to one of the sixteen initial byte values. See Fig.~\ref{fig:mula} for a C implementation. If we ignore loads and stores as well as control instructions,
 Muła's approach requires two \texttt{pshufb}, two \texttt{pand}, one \texttt{paddb}, and one \texttt{psrlw} instruction, so six inexpensive instructions to compute the population counts of sixteen bytes.
 The Muła algorithm requires fewer instructions than the part of
Wilkes-Wheel-Gill that does the same work (see Fig.~\ref{fig:wwg}), but works on twice as many
input bytes per iteration.

 \begin{figure}[h]\centering
\begin{tabular}{c}
\begin{lstlisting}
__m128i count_bytes(__m128i v) {
  __m128i lookup = _mm_setr_epi8(0,1,1,2,1,2,2,3,1,2,2,3,2,3,3,4);
  __m128i low_mask = _mm_set1_epi8(0x0f);
  __m128i lo = _mm_and_si128(v, low_mask);
  __m128i hi = _mm_and_si128(
       _mm_srli_epi16(v, 4), low_mask);
  __m128i cnt1 =
       _mm_shuffle_epi8(lookup, lo);
  __m128i cnt2 =
       _mm_shuffle_epi8(lookup, hi);
  return _mm_add_epi8(cnt1, cnt2);
}
\end{lstlisting}
\end{tabular}
\caption{\label{fig:mula}A C function using SSE intrinsics implementing Muła's algorithm to compute sixteen population counts, corresponding to sixteen input bytes. }
\end{figure}

The \texttt{count_bytes} function from Fig.~\ref{fig:mula} separately computes the population count for each of the sixteen input bytes, storing each in a separate byte of the result.   As each of these bytes will be in $[0,8]$, we can sum the result vectors from up to 31~calls to \texttt{count_bytes} using the \texttt{_mm_add_epi8} intrinsic without risk of overflow before using the \texttt{psadbw} instruction (using the \texttt{_mm_sad_epu8} intrinsic) to horizontally sum the individual bytes into two 64-bit counters.   In our implementation, we found it adequate to call the \texttt{count_bytes function} eight times between each call to \texttt{psadbw}.

Morancho observed that we can use both a vector approach, like Muła's, and the \texttt{popcnt} in a hybrid approach~\cite{6787256}. Morancho proposed a family of hybrid schemes that could be up to 22\% faster than an implementation based on \texttt{popcnt} for sufficiently large input arrays.

\begin{table*}
\caption{Relevant SSE instructions  with latencies and reciprocal throughput in CPU cycles on recent (Haswell) Intel processors \label{ref:sseinstructions}.}\centering\small
\begin{tabular}{llp{2in}cc}
\toprule
instruction & C intrinsic & description & latency &  \multicolumn{1}{p{0.7in}}{rec.\ throughput}
\\\midrule
\texttt{paddb} & \texttt{_mm_add_epi8}  & add sixteen pairs of 8-bit integers& 1 & 0.5\\
\texttt{pshufb} & \texttt{_mm_shuffle_epi8} & \emph{shuffle}  sixteen bytes & 1 & 1\\
\texttt{psrlw} & \texttt{_mm_srli_epi16} & shift right eight 16-bit integers & 1 & 1\\
\texttt{pand} & \texttt{_mm_and_si128} & 128-bit AND & 1 & 0.33\\
\texttt{psadbw} & \texttt{_mm_sad_epu8} & sum of the absolute differences of the byte values to the low 16 bits of each 64-bit word & 5 & 1\\
\bottomrule
\end{tabular}
\end{table*}

\section{Novel Vectorized Algorithms}

Starting with the Haswell microarchiture released in 2013, Intel processors support the AVX2 instruction set with 256-bit
vectors, instead of the shorter 128-bit vectors. It supports instructions and intrinsics that are analogous to the SSE intrinsics (see Table~\ref{ref:sseinstructions}).\footnote{To our knowledge, Muła was first to document the  benefits of AVX2 for the population count problem in March 2016~\cite{Mula2008}.}

The Muła function provides an effective approach
to compute population counts at a speed close to an x64 processor's
\texttt{popcnt} instruction when using 128-bit vectors, but after upgrading to AVX2's 256-bit vectors, it becomes faster than functions using the \texttt{popcnt} instruction. We present the basis of such a function in Fig.~\ref{fig:avxmula} using AVX2 intrinsics; the AVX2 intrinsics are analogous to the SSE intrinsics (see Fig.~\ref{ref:sseinstructions}). It returns a 256-bit word that can be interpreted as four 64-bit counts (each having value in $[0,64]$).
We can then add the result of repeated calls with the
\texttt{_mm256_add_epi64} intrinsic to sum 64-bit counts.

\begin{figure}[h]\centering
\begin{tabular}{c}
\begin{lstlisting}
__m256i count(__m256i v) {
  __m256i lookup =
   _mm256_setr_epi8(0, 1, 1, 2, 1, 2, 2, 3, 1, 2, 2, 3, 2, 3, 3, 4, 0, 1, 1, 2, 1, 2, 2, 3, 1, 2, 2, 3, 2, 3, 3, 4);
  __m256i low_mask = _mm256_set1_epi8(0x0f);
  __m256i lo =  = _mm256_and_si256(v, low_mask);
  __m256i hi = _mm256_and_si256(_mm256_srli_epi32(v, 4), low_mask);
  __m256i popcnt1 = _mm256_shuffle_epi8(lookup, lo);
  __m256i popcnt2 = _mm256_shuffle_epi8(lookup, hi);
  __m256i total = _mm256_add_epi8(popcnt1,popcnt2);
  return _mm256_sad_epu8(total, _mm256_setzero_si256());
}
\end{lstlisting}
\end{tabular}
\caption{\label{fig:avxmula}A C function using AVX2 intrinsics implementing Muła's algorithm to compute the four population counts of the four 64-bit words in a 256-bit vector. The 32\,B output vector should be interpreted as four~separate 64-bit counts that need to be summed to obtain the final population count.}
\end{figure}

For a slight gain in performance, we can call the Muła  function  several times while skipping the call to \texttt{_mm256_sad_epu8}, adding the byte values with \texttt{_mm256_add_epi8} before calling
\texttt{_mm256_sad_epu8} once. Each time we call
the Muła  function, we process 32~input bytes and  get 32~byte values in $[0,8]$.
We can add sixteen totals before calling \texttt{_mm256_sad_epu8} to sum the
results into four 64-bit words (since $8\times 16 =128 < 2^8$), thus processing a block of 512~bytes per call.\footnote{We could call the Muła  function up to 31~times, since $8 \times 31 = 248 < 2^{8}$.}

Of all the non-vectorized (or \emph{scalar}) functions, Harley-Seal approaches
are  fastest. Thus we were motivated to port the approach to AVX2.
The carry-save adder that worked on 64-bit words (see Fig.~\ref{fig:csa}) can be adapted in a straight-forward manner to work with  AVX2 intrinsics (see Fig.~\ref{fig:avxcsa}).

\begin{figure}[h]\centering
\begin{tabular}{c}
\begin{lstlisting}
void CSA(__m256i* h, __m256i* l, __m256i a, __m256i b, __m256i c) {
  __m256i u = _mm256_xor_si256(a , b);
  *h = _mm256_or_si256(_mm256_and_si256(a , b) , _mm256_and_si256(u , c) );
  *l = _mm256_xor_si256(u , c);
}
\end{lstlisting}
\end{tabular}
\caption{\label{fig:avxcsa}A C function using AVX2 intrinsics implementing a bitwise parallel carry-save adder (CSA).}
\end{figure}

Fig.~\ref{fig:avxhs} presents an efficient Harley-Seal function using an AVX2 carry-save adder. The function processes the data in blocks of sixteen~256-bit vectors (512\,B). It calls  Muła's AVX2 function (see Fig.~\ref{fig:avxmula}).

\begin{figure}[htb]\centering
\begin{tabular}{c}
\begin{lstlisting}
uint64_t avx_hs(__m256i* d, uint64_t size) {
 __m256i total   = _mm256_setzero_si256();
 __m256i ones   = _mm256_setzero_si256();
 __m256i twos   = _mm256_setzero_si256();
 __m256i fours   = _mm256_setzero_si256();
 __m256i eights  = _mm256_setzero_si256();
 __m256i sixteens = _mm256_setzero_si256();
 __m256i twosA, twosB, foursA, foursB,
   eightsA, eightsB;
 for(uint64_t i = 0; i < size; i += 16) {
  CSA(&twosA, &ones, ones, d[i], d[i+1]);
  CSA(&twosB, &ones, ones, d[i+2], d[i+3]);
  CSA(&foursA, &twos, twos, twosA, twosB);
  CSA(&twosA, &ones, ones, d[i+4], d[i+5]);
  CSA(&twosB, &ones, ones, d[i+6], d[i+7]);
  CSA(&foursB,& twos, twos, twosA, twosB);
  CSA(&eightsA,&fours, fours, foursA,foursB);
  CSA(&twosA, &ones, ones, d[i+8], d[i+9]);
  CSA(&twosB, &ones, ones, d[i+10],d[i+11]);
  CSA(&foursA, &twos, twos, twosA, twosB);
  CSA(&twosA, &ones, ones, d[i+12],d[i+13]);
  CSA(&twosB, &ones, ones, d[i+14],d[i+15]);
  CSA(&foursB, &twos, twos, twosA, twosB);
  CSA(&eightsB, &fours, fours, foursA, foursB);
  CSA(&sixteens, &eights, eights, eightsA, eightsB);
  total=_mm256_add_epi64(total, count(sixteens));
 }
 total = _mm256_slli_epi64(total, 4);
 total = _mm256_add_epi64(total,
   _mm256_slli_epi64(count(eights), 3));
 total = _mm256_add_epi64(total,
   _mm256_slli_epi64(count(fours), 2));
 total = _mm256_add_epi64(total,
   _mm256_slli_epi64(count(twos),  1));
 total =_mm256_add_epi64(total,count(ones));
 return _mm256_extract_epi64(total, 0)
   + _mm256_extract_epi64(total, 1)
   + _mm256_extract_epi64(total, 2)
   + _mm256_extract_epi64(total, 3);
}
\end{lstlisting}
\end{tabular}
\caption{\label{fig:avxhs}A C function using AVX2 intrinsics implementing Harley-Seal's  algorithm. It assumes, for simplicity, that the input size in 256-bit vectors is divisible by 16. See Fig.~\ref{fig:avxmula} for the \texttt{count} function.}
\end{figure}

Processors execute complex machine instructions using low-level instructions called \microops{}.
\begin{itemize}
\item Using the dedicated \texttt{popcnt} instruction for the population of an array of words requires loading the word (\texttt{movq}), counting the bits (\texttt{popcnt}), and then adding the result to the total (\texttt{addq}). The load and the \texttt{popcnt} can be combined into a single assembly instruction, but internally they are executed as separate µops, and thus each 64-bit word requires three \microops{}.  Apart from minimal loop overhead, these three
operations can be executed in a single cycle on a modern x64 superscalar
processor, for a throughput of just over one cycle per 8\,B word.
\item The AVX2 Harley-Seal function processes sixteen 256-bit vectors (512\,B) with 98~\microops{}:  16~loads (\texttt{vpmov}), 32~bitwise ANDs (\texttt{vpand}), 15~bitwise ORs (\texttt{vpor}), and 30~bitwise XORs (\texttt{vpxor}).   Each 64-bit word (8\,B) thus takes just over 1.5~\microops{}---about half as many as required to
use the builtin \texttt{popcnt} instruction on the same input.
\end{itemize}
  While fewer~\microops{} does does not guarantee faster execution, for computationally
intensive tasks such as this it often proves to be a significant
advantage.  In this case, we find that it does in fact result in
approximately twice the speed.

\section{Beyond population counts}

In practice, we often want to compute population counts on the result of some operations. For example, given two bitsets, we might want to determine the cardinality of their intersection (computed as the bit-wise logical AND) or the cardinality of their union (computed as the bit-wise logical OR). In such instances, we need to load input bytes from the two bitsets, generate a temporary word, process it to determine its population count, and so forth. When computing the Jaccard index, given that we have no prior knowledge of the population counts, we need to compute both the intersection and the union, and then we need to compute the two corresponding population counts (see Fig.~\ref{fig:popcntjaccard}).


\begin{figure}[h]\centering
\begin{tabular}{c}
\begin{lstlisting}
void popcnt_jaccard_index(uint64_t* A, uint64_t* B, size_t n) {
  double s = 0;
  double i = 0;
  for(size_t k = 0; k < n; k++) {
    s += _mm_popcnt_u64(A[k] | B[k]);
    i += _mm_popcnt_u64(A[k] & B[k]);
   }
  return i / s;
}
\end{lstlisting}
\end{tabular}
\caption{\label{fig:popcntjaccard}A C function using the \texttt{_mm_popcnt_u64} intrinsic to compute the Jaccard index of a pair of 64-bit inputs.}
\end{figure}

Both loads and logical operations benefit greatly from
vectorization, and hybrid scalar/vector approaches can be difficult
because inserting and extracting elements into and from vectors adds overhead. With AVX2, in one operation, we can load a 256-bit register or compute the logical AND between two 256-bit registers. This is four times the performance of the corresponding 64-bit operations. Thus we can expect good results from  fast population count functions based on AVX2 adapted for the computation of the Jaccard index, the cardinality of the intersection or union, or similar operations.

\section{Experimental Results}

We implemented our software in C.
We use a Linux server with an Intel i7-4770 processor running at \SI{3.4}{GHz}.
This Haswell processor has \SI{32}{kB} of L1 cache and \SI{256}{kB} of L2 cache per core with \SI{8}{MB} of L3 cache.
The machine has \SI{32}{GB} of RAM (DDR3-1600 with double-channel). We disabled Turbo Boost and set the processor
to run at its highest clock speed. Our software is freely available
 (\url{https://github.com/CountOnes/hamming_weight})
and was  compiled using the GNU GCC~5.3 compiler with  the ``\texttt{-O3 -march=native}'' flags.

For all experiments, we use randomized input bits. However, we find that the performance results are insensitive to the data values.

Table~\ref{tab:speed} presents our results in number of cycles per word, for single-threaded execution. To make sure our results are reliable, we repeat each test 500~times and check that the minimum and the average cycle counts are within 1\% of each other.
We report the minimum cycle count divided
by the number of words in the input. All the scalar methods
(WWG, Laradoux, and HS) are significantly slower than the native
\texttt{popcnt}-based function.
We omit tabulation-based approaches from  Table~\ref{tab:speed} because they are not competitive: 16-bit tabulation uses over 5~cycles
even for large arrays.
We can see that for inputs larger than 4\,kB, the AVX2-based Harley-Seal approach is
twice as fast as our optimized \texttt{popcnt}-based function, while for small arrays (fewer 64~words) the \texttt{popcnt}-based function is fastest.

\begin{table*}
\caption{\label{tab:speed}Number of cycles per 64-bit input word required to compute the population of arrays of various sizes.}\centering\small
\begin{tabular}{c|cccccc}
\toprule
array size & WWG & Lauradoux & HS & \texttt{popcnt} & AVX2 Muła & AVX2 HS\\
\\\midrule
256\,B& 6.00 & 4.50 &3.25 & \textbf{1.12} & 1.38 & ---\\
512\,B& 5.56 & 2.88 &2.88 & 1.06 & \textbf{0.94} & ---\\
1\,kB& 5.38 & 3.62 & 2.66 & 1.03 & 0.81 & \textbf{0.69}\\
2\,kB& 5.30 & 3.45 & 2.55 & 1.01 & 0.73 & \textbf{0.61}\\
4\,kB& 5.24 & 3.41 & 2.53 & 1.01 & 0.70 & \textbf{0.54}\\
8\,kB& 5.24 & 3.36 & 2.42 & 1.01 & 0.69 & \textbf{0.52}\\
16\,kB& 5.22 & 3.36 & 2.40 & 1.01 & 0.69 & \textbf{0.52}\\
32\,kB& 5.23 & 3.34 & 2.40 & 1.01 & 0.69 & \textbf{0.52}\\
64\,kB& 5.22 & 3.34 & 2.40 & 1.01 & 0.69 & \textbf{0.52}\\
\bottomrule
\end{tabular}
\end{table*}

We present the results for Jaccard index computations in Table~\ref{tab:speedji}. Contrary to straight population counts, the Jaccard-index AVX2 Muła remains faster than the \texttt{popcnt}-based function even for small blocks (256\,B). 
AVX2 HS provides the best speed, requiring only 1.15~cycles to
calculate the Jaccard similarity between each pair of 64-bit inputs.
This is more than twice as fast ($2.4\times$) as the \texttt{popcnt}-based function.   Since
the population count is done for each input of the pair, the speed of the similarity is only slightly greater than the speed of calculating the two population counts individually.  That
is, using AVX2 for both the Boolean operation and both population
counts gives us the Boolean operation almost for free.

\begin{table}
\caption{\label{tab:speedji}Number of cycles per pair of 64-bit input words required to compute the Jaccard index of arrays of various sizes.}\centering\small
\begin{tabular}{c|ccc}
\toprule
array size &  \texttt{popcnt} & AVX2 Muła & AVX2 HS\\
\\\midrule
256\,B & 3.00 & \textbf{2.50} & ---\\
512\,B & 2.88 & \textbf{2.00} & ---\\
1\,kB&  2.94 & 2.00 & \textbf{1.53}\\
2\,kB&  2.83 & 1.84 & \textbf{1.33}\\
4\,kB&  2.80 & 1.76 & \textbf{1.22}\\
8\,kB&  2.78 & 1.75 & \textbf{1.16}\\
16\,kB& 2.77 & 1.75 & \textbf{1.15}\\
32\,kB& 2.76 & 1.75 & \textbf{1.15}\\
64\,kB& 2.76 & 1.74 & \textbf{1.15}\\
\bottomrule
\end{tabular}
\end{table}

\section{Conclusion}

On recent Intel processors, the fastest approach to compute the population count on moderately large arrays (e.g., 4\,kB) relies on a vectorized version of the Harley-Seal  function. It is twice as fast as functions based on the dedicated instruction (\texttt{popcnt}). For the computation of similarity functions between two bitsets, a vectorized approach based on the Harley-Seal function is more than twice as fast ($2.4\times$) as an optimized approach based on the \texttt{popcnt} instruction.

Future work should consider other computer architectures. Building on our work~\cite{Mula2008}, Sun and del Mundo tested various population-count functions on an Nvidia GPU and found that its popcount  intrinsic gave the best results~\cite{sunrevisiting}.

\ack
We are grateful to S.~Utcke for pointing out a typographical error in one of the figures.


\section*{Funding}
This work was supported by  the Natural Sciences and Engineering Research Council of Canada [261437].

\bibliographystyle{compj}
\bibliography{hamming}

\appendix

\section{Population Counts in AVX-512}

Though Intel processors through the current Kaby Lake generation do not
yet support the AVX-512 instruction set, it is straight-forward to generalize our vectorized algorithms to 512-bit vectors.
However, even beyond the increase in vector width, it should be
possible to implement the carry-save adder more efficiently with
AVX-512, which also adds the \texttt{vpternlogd} instruction.   Available
through C intrinsics as \texttt{_mm512_ternarylogic_epi64}, this instruction
allows us to compute arbitrary three-input binary functions in a
single operation.  Utilizing this, we can replace the 5 logical
instructions we needed for AVX2 with just two instructions.
The \texttt{vpternlogd} instruction relies on an integer parameter $i$ that serves as a look-up table. Given the input bits $x,y,z$, the value given that the $(x+2y+4z)^{\mathrm{th}}$ bit of the parameter $i$ is returned. For example, to compute XOR of the inputs,  the $i$ parameter needs to have a 1-bit at indexes 1, 2, 4 and 7 (i.e., $i=2^1+2^2+2^4+2^7=150$ or \texttt{0x96} in hexadecimal). Similarly, to compute the most significant bit of the carry-save adder, the $i$ parameter needs to have a 1-bit at indexes 3, 5, 6 and 7 (\texttt{0xe8} in hexadecimal). Fig.~\ref{fig:avx512csa} presents a C function implementing a carry-save adder (CSA) using AVX-512 intrinsics.

\begin{figure}[bht]\centering
\begin{tabular}{c}
\begin{lstlisting}
void CSA(__m512i* h, __m512i* l, __m512i a, __m512i b, __m512i c) {
  *l = _mm512_ternarylogic_epi32(c, b, a, 0x96);
  *h = _mm512_ternarylogic_epi32(c, b, a, 0xe8);
}
\end{lstlisting}
\end{tabular}
\caption{\label{fig:avx512csa}A C function using AVX-512 intrinsics implementing a bitwise parallel carry-save adder (CSA).}
\end{figure}

Intel has also announced that future processors might support the AVX-512 \texttt{vpopcnt} instruction~\cite{intel2016} which computes the population count of each word in a vector. To our knowledge, no available processor supports \texttt{vpopcnt} at this time.

\end{document}